\documentclass[twocolumn,prl,aps,amsmath,showpacs]{revtex4}

\usepackage{graphics}

\usepackage{dcolumn}

\usepackage{bm}

\usepackage{epsfig}

\begin{document}

\title{Metal-insulator transition
and charge ordering in the extended\\
 Hubbard model at one-quarter filling}
\author{M. Calandra}
\email[]{calandra@lmcp.jussieu.fr}    
\altaffiliation[Present address: ] {Laboratoire de Min\'eralogie-cristallographie, 
Universit\'e Pierre et Marie curie, 
4 Place Jussieu, 
75252 Paris, France} 
\author{J. Merino}
\affiliation{Max-Planck-Institut f\"ur Festk\"orperforschung
D-70506 Stuttgart, Germany}
\author{Ross H. McKenzie}
\affiliation{Department of Physics, University of Queensland,
Brisbane 4072,
Australia}
\date{today}

\begin{abstract}
We study with exact diagonalization the zero temperature
properties of the quarter-filled extended Hubbard model on a
square lattice. We find that increasing the ratio of the intersite Coulomb
repulsion, $V$, to the band width
drives the system from a metal to a charge ordered
insulator.
 The evolution of the optical conductivity spectrum
with increasing $V$ is compared to  the observed optical 
conductivity of several layered molecular crystals with the
$\theta$ and $\beta''$ crystal structures.
\end{abstract}

\maketitle

Charge ordering in 
strongly correlated electron systems   is  
currently under intense investigation. Charge ordering is    
relevant  to a broad range of materials including the
cuprates\cite{cuprate},
manganates\cite{cmr}, magnetite\cite{mag},
vanadium oxides\cite{van}, and the Bechgaard
salts  
\cite{brown}.
 The $\theta$ and $\beta''$ types of layered
molecular crystals based on molecules such as BEDT-TTF  [=
bisethylenedithio-tetrathiafulvalene] \cite{Ishiguro}, 
display charge ordering, metallic, and superconducting phases close
to each other \cite{mi}.
 Charge ordering driven by a   strong
inter-site Coulomb repulsion\cite{hubbard,Seo}  is possible 
in crystals with the $\theta$ and
$\beta''$ arrangements of BEDT-TTF molecules because  
their bands are quarter-filled with holes, in contrast to 
the well studied $\kappa$-type, for which strong
dimerization of the molecules lead to a half-filled band \cite{kino}.  
The $\theta$-type crystals
undergo a transition from a metal to a charge ordered
insulator as the temperature, pressure, uniaxial stress,
or anion is varied\cite{mi,hubbard}.
Furthermore, the metallic phase exhibits features
characteristic of a strongly correlated system.
In particular, the optical conductivity spectra 
display a broad mid-infrared band and a near absence
of a Drude-like peak\cite{Dong}. This is in contrast to conventional metals,
for which the total spectral weight is dominated by
a Drude peak.

In this Letter, we use the
results of an exact diagonalization
study of the relevant extended Hubbard
model to argue that
the inter-site Coulomb repulsion 
is responsible for the observed
metal-insulator transition in the $\theta$ and $\beta''$ crystals.
 We show how the Drude
weight decreases as the inter-site Coulomb repulsion, $V$, is
increased, until, at a finite value of $V$,  a transition 
to an insulating phase occurs. Simultaneously, long range charge 
ordering gradually sets in. We further find that a redistribution of
the optical conductivity spectra occurs close to the
metal-insulator transition.
This finding is in qualitative agreement with
experimental data on $\theta$ and $\beta''$ organic salts
\cite{Tajima,Wang,Dong,Ouyang}.

The quarter-filled extended Hubbard model on a square lattice is
the {\it simplest} strongly correlated model that can potentially
describe the competition between metallic, superconducting,
and insulating phases in the 
$\theta$ and $\beta''$ materials \cite{McKenzie,merino}.
The Hamiltonian is 
\begin{eqnarray}
H = &-t& \sum_{<ij>,\sigma} (c^\dagger_{i \sigma} c_{j \sigma} +
c^\dagger_{j \sigma} c_{i \sigma}) + U \sum_{i} n_{i\uparrow}
n_{i\downarrow}
\nonumber \\
&+& V \sum_{<ij>} n_i n_j
\label{ham}
\end{eqnarray}
where $c^\dagger_{i \sigma}$ creates an electron of spin $\sigma$
at site $i$.
For $V = 0$, previous calculations suggest
that the system is metallic with no
charge order \cite{DagottoRMP}. In the limit of $U, V 
\gg t$  the  double occupation of
sites is suppressed and the ground state is insulating with
checkerboard charge ordering and long range antiferromagnetic
correlations along the diagonals \cite{McKenzie}.

In the present work we consider large (but finite) $U/t$ and vary
$V/t$ \cite{parameters}. 
Previously, an SU(N) generalisation of the model
with $U \to \infty$,
was studied in the large N limit using slave bosons\cite{McKenzie,merino}.
It was found that as $V/t$ increased there was a transition
from a metallic phase to a superconducting phase
(with $d_{xy}$ symmetry) to a charge ordered phase.
Given this potentially rich phase diagram it is important 
to determine
whether or not these previous results are an artefact
of the approximations used or whether they reflect the
actual physics (N=2 and finite $U/t$). Exact diagonalization of
small systems provides such a test.
Previously, similar models
with more parameters, aimed at a more
realistic description of the details of
the materials, have been studied by
Hartree-Fock\cite{Seo}, and quantum Monte Carlo\cite{clay1},
and exact diagonalization\cite{Clay}. 
We performed a Lanczos calculation on $L=8, 16, 20$ site
clusters\cite{DagottoRMP} avoiding uncompensated spin moments in
the cluster ($N_{\uparrow}=N_{\downarrow}$ ) at quarter-filling:
$<n>=1/2$.
A powerful method
to determine from small systems
whether the bulk system is metallic or insulating
is to evaluate the Drude weight $D$ \cite{Kohn,DagottotU,Scalapino}.
It is given by
\begin{equation}
\frac{D}{2 \pi e^2} = - \frac {\langle 0|T|0\rangle}{4 L}  - \frac{1}
{L} \sum_{n \ne 0} \frac{|\langle n|j_x|0\rangle|^2}{E_n-E_0}
\label{drude}
\end{equation}
where $E_0$ and $E_n$, denote the ground state and excited state energies
of the system, respectively. $T$ is the kinetic energy operator
and $j_x$ is the current operator
 in the $x$ direction
 at zero wavevector (${\bf q}=0$).
 The occurrence of an insulating phase is marked by the
exponential vanishing of $D$ with the linear size of the system
$L_x = \sqrt{L}$ \cite{Kohn,Scalapino,chargeg}.

\begin{figure}
\resizebox{!}{2.5in}{\includegraphics{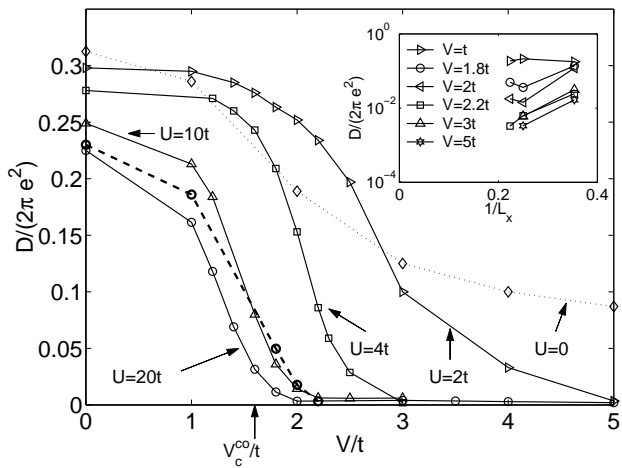}}
\caption[]{Metal-insulator
transition induced by the nearest-neighbour
Coulomb interaction $V$.
The Drude weight, $D$, is shown as a function of  $V/t$, for $L=16$
and various values of $U/t$
(continous and dotted lines) and for $L=20$, $U=10t$ (dashed line).
The arrow in the horizontal axis marks the
onset of checkerboard charge ordering at $V=V_c^{CO}$, for $U=10
t$ (see Fig. \ref{fig2}).  The inset shows
the  finite-size scaling of $D$ as a function
of $1/L_x$ for different values of $V/t$ with $U=10t$. 
The metal-insulator transition occurs at 
$V_c^{MI} \approx 2.2t$.} 
\label{fig1}
\end{figure}

In Fig. \ref{fig1} we plot the Drude weight as a function of $V/t$,
for $L=16$ and different values of $U/t$.
For $U=10 t$ we also plot
the Drude weight for the cluster with $L=20$. As the Coulomb
nearest-neighbors repulsion is gradually increased, the Drude
weight decreases until it eventually vanishes. For $L=16$ we
estimate the critical value for the metal-insulator transition to
be $V_c^{MI} \approx 2.2 t$; for this value, the Drude weight
becomes of the order of $10^{-3}$ and stays of this order up to
$V=5 t$. For $L=20$ we do not find any significant change in the
dependence of the Drude weight with $V$, meaning that the finite size
effects in the Drude weight are weak.
This is confirmed in the
inset of Fig. \ref{fig1} which shows the finite-size scaling of the
Drude weight. Indeed, we find that for $V > V_c^{MI} \approx 2.2 t
$, the Drude weight displays an exponential dependence with
$1/L_x$, as expected for an insulator\cite{Kohn,Scalapino}.
It extrapolates to values of the order of $10^{-5}$ in
the thermodynamic limit. In contrast, for $V < V_c^{MI}$, the
Drude weight is weakly dependent on               $1/L_x$
extrapolating to a finite value in the thermodynamic limit,
consistent with a metallic state.

\begin{figure}
\resizebox{!}{2.4 in}{\includegraphics{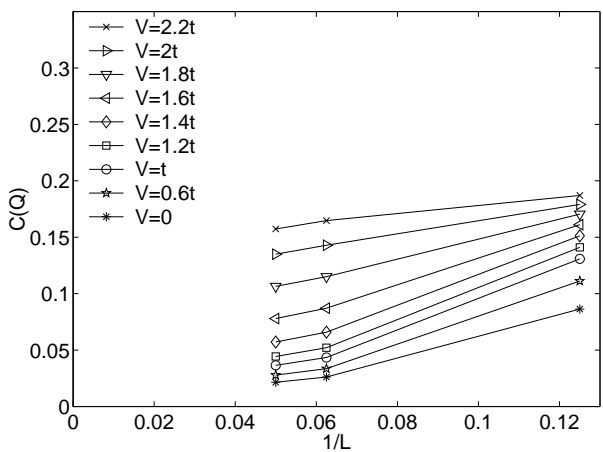}}
\caption{Checkerboard charge ordering induced by
the nearest neighbour
Coulomb interaction.
Finite-size scaling of the charge correlation
function $C({\bf q})$ (defined in Eqn. (\ref{charge})) 
at ${\bf q}={\bf Q} \equiv(\pi,\pi)$
for $U=10t$ and several values of $V/t$ at
quarter filling.  Long range checkerboard charge ordering occurs
for $V>V_c^{CO}\approx 1.6 t$. } \label{fig2}.
\end{figure}

The occurrence of charge ordering can be investigated by computing
the charge correlation function
\begin{equation}
C({\bf q})= \frac{1}{L} \sum_{ij} e^{i {\bf q} \cdot ({\bf R_i-R_j})}
<n_i n_j> \label{charge}
\end{equation}
where $ {\bf q}$ are the allowed momenta on the cluster. We find
that at $V=0$, $C({\bf q})$ is featureless. As $V$ is increased,
$C({\bf q})$ peaks at ${\bf q}={\bf Q} \equiv(\pi,\pi)$,
 signalling checkerboard
charge ordering.
In our calculations we do not find any evidence  for other charge
ordering patterns, 
such as those found in References  \cite{Seo,Clay}.
The system has long range charge ordering if $C({\bf Q})$ remains
finite in the thermodynamic limit. In Fig. \ref{fig2} we plot
$C({\bf Q})$ as a function of $1/L$.
A linear dependence with
$1/L$ is expected for the finite size scaling of an order
parameter with a discrete symmetry in 
two dimensions\cite{Brezin}. For $V=0$,
we observe that the linear extrapolation of
$C({\bf Q})$ (using $L=16, 20$) to the
infinite volume limit tends to zero, {\it i.e.}, the charge is
homogeneously distributed in the lattice. 
 We find
that the extrapolation of a linear fit of $L=8, 16, 20$ to the
thermodynamic limit becomes finite at about
 $V_c^{CO} \approx 1.6 t$ 
\cite{extrapol}.
 If we decrease $U$ the critical value
$V_c^{CO}$ increases. This is because doubly-occupied sites
proliferate so that inducing the checkerboard charge ordered state
becomes energetically less favorable.
Further insight into the transition can be obtained by computing the
charge ordering parameter\cite{Vojta}, $\eta=\sqrt{C({\bf
Q})/<n^2>}$. A non-zero $\eta$ means long range charge
ordering. In particular, for $\eta=1$ the checkerboard charge
ordered state is fully formed, while for $\eta=0$ the charge is
homogeneously distributed in the lattice. At $V = 2.2 t$ we find
that the checkerboard is close to being completed, $\eta \approx 0.75$. 
We note that the results we find here are different from the ones obtained in
two-leg ladders \cite{Vojta} where a transition from an
homogeneous insulating phase to a charge ordered insulating phase
takes place. This difference is     due to the one-dimensional nature of the
two-leg ladder.

At this stage we can provide a physical interpretation of our
findings. Only when $V$ is sufficiently large so that the
checkerboard charge ordered state is nearly completed, a
transition to an insulating state is possible. This is because
moving an electron within the checkerboard ordered state
would cost an energy of $\approx 3 V $, which is comparable to  the
bandwidth, $W=8 t$. As a consequence electrons can become
localized. 
As an aside we note that
 Fig. \ref{fig1} shows that increasing $U$ from $10
t$ to $20 t$, leads to only a slight decrease
in the   critical value,
$V_c^{MI}$. This contradicts the results
obtained by Ohta {\it et al.} \cite{Ohta} who claim that $V_c
\approx U/4$. We note that the critical value $V_c^{MI}$ for large
$U$ follows the tendency encountered in one-dimensional rings
\cite{Mila} and two-leg ladders \cite{Vojta} for which $V_c^{MI}$
decreases to $2 t$ as $U \rightarrow \infty $.

Since we find that $V_c^{MI} > V_c^{CO}$, it might be possible
that a charge ordered metallic state is realized in the range:
$V_c^{CO}< V < V_c^{MI}$. 
However, caution is in order because
the former value is quite sensitive to finite-size effects.
A possible candidate for this phase is
the quarter-filled organic crystal:
 $\beta$''-(BEDT-TTF)$_2$SF$_5$CH$_2$CF$_2$SO$_3$,
 which displays metallic behavior of
the resistivity and at the same time charge disproportion in
alternate molecules\cite{Schlueter}.

In order to further our understanding of the metal-insulator
transition and make contact with experiments on $\theta$ and
$\beta''$  materials, we have also
computed the real part,
$\sigma (\omega)$,
 of the optical conductivity\cite{favand}
 at frequency $\omega$,
\begin{equation}
 \sigma (\omega)=D \delta(\omega) + \frac{\pi e^2}{L} \sum_{n \ne
0}\frac{|\langle n|j_x|0\rangle|^2}{E_n-E_0} \delta( \omega -
E_n +E_0).
\label{opt}
\end{equation}
It obeys the following sum rule\cite{Maldague}
\begin{equation}
\int_0^{\infty}  \sigma (\omega) d \omega =  - \frac{\pi e^2}{2 L}
<0| T |0>.
\label{sum}
\end{equation}

\begin{figure}
\resizebox{!}{2.5in}{\includegraphics{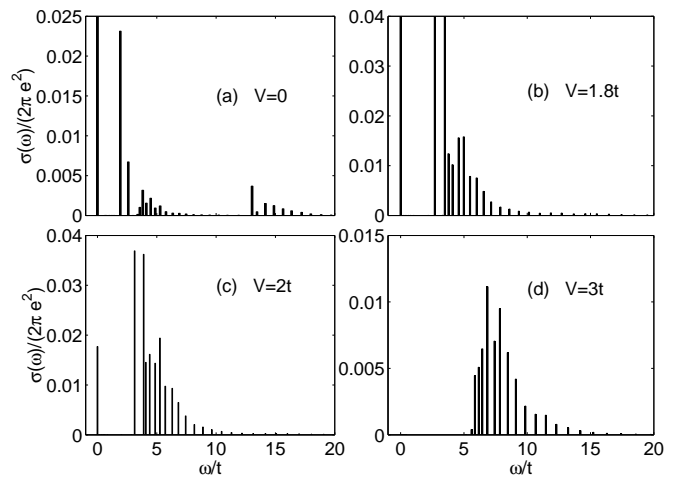}}
\caption{Transfer of spectral weight in
the optical conductivity with increasing
$V/t$. The frequency dependence of
the real part of the  conductivity
$\sigma(\omega)$ for $L=20$ and $U=10t$ is
shown in both the metallic 
[$V=0$ (a), $V= 1.8 t$ (b), $V= 2 t$ (c)],
and insulating  [$V=3t$ (d)] phases.
Note that close to the metal-insulator transition
the spectral weight from the Drude peak and the Hubbard
feature   is transferred to the ``mid-infrared'' peak.}
\label{fig3}
\end{figure}

In Fig. \ref{fig3} we plot the evolution of the optical
conductivity for different values of $V/t$ and $L=20$. We also found
that
the plots were qualitatively similar for $L=16$ (not shown). For
the regular Hubbard model, {\it i.e.} $V=0$ (Fig. \ref{fig3} (a)), there
are two features at non-zero frequency.
 The highest energy feature is related to transitions associated with the
Hubbard bands and the other is the ``mid-infrared'' band, previously
found by Dagotto {\it et al.}\cite{DagottotU}.
 At $V= 1.8 t$ (Fig.  \ref{fig3} (b)),
 the Drude peak is reduced (see Fig. \ref{fig1}), signalling the incipient
localization of the charge carriers. Spectral weight is
transferred from both the Hubbard feature and the
Drude peak to the mid-infrared band found at $V=0$. This effect is more clearly
observed for $V=2 t$ in Fig. \ref{fig3} (c):
the Drude peak is further suppressed and most of the spectral weight
appears at low frequencies, in the range $V < \omega < 3V$. Increasing $V$
further to just below $ V_{MI} $
gives a similar shape of the optical conductivity (not
shown) with a very small Drude weight. 
Finally, Fig. \ref{fig3} (d) shows the optical
conductivity on the insulating side of the transition: $V=3 t$. An
optical gap opens and most of the spectral weight concentrates at
larger frequencies building up a single 
broad resonance. If $V$ is further increased, $V
>> U/4 $, then the band associated with charge excitations due to
$V$ appears above the Hubbard band situated near $\omega \sim U$,
 as discussed by
Ohta {\it et al.} \cite{Ohta}.
In summary, Fig. 3 shows that close to the metal-insulator
transition the optical conductivity is dominated
by the "mid-infrared" feature, for both metallic
and insulating phases. We now discuss 
how this is what is observed in the
$\theta$ and $\beta''$ materials.

The optical conductivity we find close to 
the metal-insulator transition, $V \alt V_c^{MI}$, 
is qualitatively similar to the
optical conductivity of
 $\beta''$-(BEDT-TTF)$_2$SF$_5$CH$_2$CF$_2$SO$_3$,
 measured by Dong {\it et al.} \cite{Dong}
 (at a temperature, $T=14$ K). 
 Indeed, the small spectral
weight at zero frequency found here is consistent with their
failure to observe a 
Drude peak, even though the system is metallic.
 Also their observation of large
spectral weight at low frequencies is consistent
with the mid-infrared band we find in our calculations.
 At $V \alt V_c^{MI} $, the system is
still on the metallic side of the metal-insulator transition which
is consistent with the observed metallic behavior of the
resistivity. From the above discussion we conclude that $\beta''$-
(BEDT-TTF)$_2$SF$_5$CH$_2$CF$_2$SO$_3$ is a 
metal close to a metal-insulator transition. This assertion is
corroborated by the fact that an external perturbation, such as
a magnetic field\cite{wosnitza1} or pressure\cite{Wosnitza}, 
or replacing the anion CH$_2$CF$_2$ by CH$_2$ \cite{Ward},
drives the system into an insulating phase.

Optical conductivity measurements on the
$\theta$-type crystals all show a broad feature
from around 1000 to 5000 cm$^{-1}$ \cite{Tamura,Tajima,Wang,Ouyang}.
We assign this to the "mid-infrared" band.
We now compare this interpretation to previous work.
$\theta$-(BEDT-TTF)$_2$CsZn(SCN)$_4$, is  a metal above
20 K and an insulator below 20 K.
The frequency-dependent conductivity from 650 to 
5000 cm$^{-1}$ at temperatures above 20 K shows
a feature from 650 to about 1200
cm$^{-1}$ \cite{Tajima}, which  Tajima {\it et al.}
assigned to a Drude peak,
 with a phenomenological damping rate of
0.1 eV $\simeq$ 800 cm$^{-1}$.
$\theta$-(BEDT-TTF)$_2$I$_3$ is a metal
which undergoes a transition to a superconductor
at 3.6 K. Tamura {\it et al.} \cite{Tamura}
measured the optical conductivity
from about 700 to 4000 cm$^{-1}$ in the metallic phase.
They assign the spectrum below about
1000 cm$^{-1}$ to the tail of a 
Drude peak with a phenomenological damping rate of
400  to 800 cm$^{-1}$, depending on the temperature.
We disagree with the assignment of these low energy
features to a Drude peak because it requires 
large scattering rates, comparable to $t$, 
implying a ``bad'' metal.
In the $\kappa$ materials the observed widths
of the Drude peak (and thus the scattering rate)
at low temperatures are orders of magnitude smaller 
than this, typically of the order of tens of
cm$^{-1}$ \cite{eldridge}.
We would assign all of the spectrum observed 
for $\theta$-(BEDT-TTF)$_2$I$_3$ 
\cite{Tamura} to the "mid-infrared" band.
The experiment of Tamura {\it et al.}\cite{Tamura} did not
go to low enough frequencies to observe the actual Drude peak.
The feature in
$\theta$-(BEDT-TTF)$_2$CsZn(SCN)$_4$ that 
Tajima {\it et al.} \cite{Tajima} assigned to a Drude peak
we would assign to a phonon (and possibly its
interference with the tail of the "mid-infrared" band)
for the following reason.
The optical conductivity spectrum of 
$\theta$-(BEDT-TTF)$_2$RbCo(SCN)$_4$
was studied in detail by
Tajima {\it et al.}\cite{Tajima}.
This material becomes a charge ordered insulator
below 190 K.
In slowly cooled samples the only
significant spectral weight below 2000 cm$^{-1}$
is a phonon around 1200 cm$^{-1}$
(compare Fig. 5(a) in Ref. \onlinecite{Tajima}).
If the sample is quenched (i.e., rapidly cooled) there 
is a broad feature around this frequency.
In the organics such quenching is usually associated
with significant amounts of disorder \cite{stalcup}.

Tajima {\it et al.} \cite{Tajima}
 assigned the second broad feature extending from
above about 1200
cm$^{-1}$ to domains associated with the onset of charge ordering.
This was based on the calculated optical
conductivity for a mean-field solution of the extended
Hubbard model.
Although we have 
essentially the same assignment as theirs
for the "mid-infrared" band we
note that 
their mean-field calculation 
cannot actually produce this feature in the metallic phase.
This underscores the many-body physics underlying
this feature and the need for our exact
diagonalization calculation.




We suggest two possible reasons for
the absence of the Hubbard resonance in
the experimental spectra.
First, Fig. 3 (c) and (d) suggest that
most of the spectral weight from it is
transfered to the "mid-infrared" band  
Second, it could be at a frequency above
the range of the experiments.
The most reliable estimates put
$U-V \simeq 0.5 $ eV $\simeq 4000$ cm$^{-1}$ \cite{parameters}.
If $V \simeq 2 t \sim 0.1 $ eV \cite{McKenzie}, Fig. 3 (c)
would put the mid-infrared band
extending from about 1000 cm$^{-1}$
to 4000 cm$^{-1}$ and the Hubbard band feature
roughly around 6000 cm$^{-1}$. The latter is
near the edge of many of the experimental
plots.

In summary, we have shown that increasing the nearest-neighbors
Coulomb repulsion, $V$, for the quarter-filled extended Hubbard
model on the square lattice, leads to 
an insulating phase for $ V > 2.2 t$.
The calculated optical conductivity spectra close to the
metal-insulator transition is consistent with experimental data on
several $\theta$ and $\beta''$- type molecular crystals.
 At $V \approx V_c^{MI}$, the Drude peak is strongly supressed
and the spectral weight is dominated by a
broad "mid-infrared" band associated
with short-range charge ordering.






\acknowledgments

We thank O. Gunnarsson for making available his exact
diagonalization code, E. Koch, M. Capone, F. Becca,
M. Dressel, and M. Brunner
for very helpful discussions, and the Max-Planck-Forschungspreis
for financial support. J.M. was supported by a Marie Curie Fellowship
of the European Community programme "Improving Human Potential" under
contract No. HPMF-CT-2000-00870 and R.H.M. 
by the Australian Research council.


\begin{thebibliography}{}
\bibitem{chargeg} We find that, unlike the Drude weight, the
charge gap: $\Delta_c=E_0(N+1)+E_0(N-1)-2 E_0(N)$, is strongly
affected by finite size effects.
\end{thebibliography}
\end{document}